# Reverse Monte Carlo and Rietveld modelling of BaMn(Fe,V)F$_7$ glass structures from neutron data

Armel Le Bail

Université du Maine, Laboratoire des Fluorures, CNRS UMR 6010, Avenue O. Messiaen, 72085 Le Mans Cedex 9, France.

E-mail address : alb@cristal.org

## Abstract

Fluoride glasses BaMnMF$_7$ (M = Fe, V, assuming isomorphous replacement) have been structurally modelled. The neutron patterns were simulated by the reverse Monte Carlo (RMC) and the Rietveld for disordered materials (RDM) methods. The best fit is obtained from one of the seven tested crystalline models : the BaMnGaF$_7$ structure-type in which the glasses crystallize. However, by RMC applied to enlarged RDM models, the difference between the best and the worst fit is quite small ($\Delta Rp \sim 0.5\%$), suggesting that two interference functions are not enough for the characterization of such multicomponent glasses, or that the average short and medium range order is similar in the seven models in spite of strong differences in the polyhedra linkage.

## Introduction

Structure simulations by the reverse Monte Carlo (RMC) method [1] applied to starting models built up from enlarged crystal structures, selected from the quality level of a Rietveld [2] fit of their scattering data, were reported for glassy SiO$_2$, ZnCl$_2$, and NaPb(Fe,V)$_2$F$_9$ [3]. The Rietveld for disordered materials (RDM) method has previously shown its potentiality to reveal very fast (in quite small computing time) if a given crystalline model would be a good starting point for further large-scale modelling by RMC [4-5]. This approach is used here for modelling the structure of barium-manganese-(iron,vanadium) fluoride glasses for a special composition BaMn(Fe,V)F$_7$ selected because it corresponds to the existence of a large number of known different crystal structures, and because of the quasi-isomorphous Fe/V substitution in fluoride materials, which is very interesting for a neutron scattering experiment.

## Experimental

The BaMn(Fe,V)F$_7$ glasses were prepared by melting the anhydrous fluoride mixtures in a dry box (inert atmosphere), then the melts, in a covered platinum crucible, were cast and rolled in a bronze mould heatted at 200°C. Density measurements provide the same number for the two glasses : $\rho_0 = 0.0710 \pm 0.0003$ atom.Å$^{-3}$. Neutron data for both BaMnFeF$_7$ and BaMnVF$_7$ glasses were recorded at ILL (Grenoble) on the D4 instrument (wavelength $\lambda = 0.497$ Å). The expected isomorphous replacement between Fe$^{3+}$ and V$^{3+}$ is well supported by the crystal chemistry in fluoride compounds. As a rule, when a $Fe^{3+}$-based crystalline fluoride exists, the isostructural equivalent $V^{3+}$ material can be prepared too, with generally no more than 1% variation in cell dimensions. Software used were ARITVE for RDM modelling, RMCA for RMC modelling, and GLASSVIR for visualization of large models in VRML (virtual reality modelling language).

**Models**

The title glasses are so-called TMFG (3d transition metal fluoride glasses) opposed to HMFG (heavy metal fluoride glasses, Zr-based). With the 3d elements generally in octahedral coordination, TMFG are considered as network glasses. Previous models for $NaPb(Fe,V)_2F_9$ glasses [4] favoured a three-dimensional (3D) organization of intercrossed octahedral chains derived from the $NaPbFe_2F_9$ crystal structure [6]. RMCA modelling with silica tetrahedral 3D networks is rather easy when using coordination and distance constraints (square plane instead of tetrahedra is avoided in principle), though obtaining systematically 4 oxygen atoms around one Si atom may ask for quite long calculation times if every O atom is not shared by two Si atoms (in such a case, a random network of Si atoms can be built first and then O atoms are inserted systematically at medium distance of two Si atoms). Octahedral 3D networks are even more difficult to build because all intermediary shapes between a trigonal prism and an octahedron can appear easily. Moreover if the tetrahedra linkage is realized by corner sharing, this is not necessarily the case of the octahedra into the $BaMn(Fe,V)F_7$ glasses. Examples are shown below presenting edge sharing in crystallized compounds. Compared to the $NaPb(Fe,V)_2F_9$ glasses, the M/F ratio changes dramatically from 2/9 to 2/7. This ratio would be 2/6 for a 3D network of corner sharing octahedra (with $MF_3$ formulation), the title glasses are very close to it. The barium atom appears however too large for being inserted into a cubic cage of eight $MF_6$ octahedra connected by corners.

Previous EXAFS results on the title glasses favour a $MF_6$ coordination [7] (M = Mn, Fe or V), with Fe-F and Mn-F mean distances being estimated respectively to 1.93(1) and 2.09(2) Å. However, if nothing else than $FeF_6$ and $VF_6$ octahedra are known in crystallized fluoride compounds (contrarily to oxides), how to be sure if $MnF_7$ or $MnF_8$ polyhedra are not existing inside of the glass, just like in some fluoride crystallized compounds ? Moreover, EXAFS remained unclear about the M-M linkage (corner or edge sharing or both). The present study by glass structure modelling would like to try to provide answers, in spite of the difficulties. Possibilities for the crystallization of $BaMM'F_7$ compounds are quite large, to a point which one may wonder if this is not one good reason for easy glass forming (and moreover, several of these crystallized compounds present systematic disorder). Seven topologically different crystal structure types appear suitable for attempting the $BaMn(Fe,V)F_7$ glass modelling. Some show edge sharing only between $MnF_6$ octahedra, or between $MnF_6$ octahedra and $MnF_8$ square antiprisms, etc. These seven structure types will be noted I-VII, they are : $BaMnFeF_7$ (I) [8], $BaMnGaF_7$ (II) [9], $BaZnFeF_7$ (III) [10], $BaCaGaF_7$ (IV) [11], $BaCuFeF_7$ (V) [12], $BaCuInF_7$ (VI) [13] and $BaNaZrF_7$ (VII) [14]. All these models were tested with the RDM method by fitting simultaneously the neutron scattering data of both Fe- and V-based glasses. Then the small crystalline cells were enlarged in order to build 4-5000 atoms models for running the RMC simulations using distance restraints and coordination constraints on the 3d elements (so that the crystalline model connectivity was kept intact). Also, fully random models were built up by the RMC method, using distance restraints and coordination constraints as well. Interestingly, a $BaMnFeF_7$ glass does not crystallize into the structure-type I. At 440°C, without particular precaution (in air) : $BaMnFeF_7$ and $BaMnVF_7$ glasses crystallize both at 100% into the $BaMnGaF_7$ structure-type II. A $BaZnGaF_7$ glass crystallizes into the BaMnFeF$_7$ structure-type (I), and a BaCuGaF$_7$ glass crystallizes into the BaCuFeF$_7$ structure-type (V). The seven models are described below.

BaMnFeF$_7$ - type I. When synthesized by heating appropriate components in platinum tubes sealed under inert atmosphere, crystals of BaMnFeF$_7$ composition belong to the monoclinic symmetry [8]. Type I consists in a three-dimensional octahedral lattice with edge-sharing dinuclear Mn$_2$F$_{10}$ units linked by corners to FeF$_6$ octahedra. But a glass with BaMnFeF$_7$ composition does not crystallize in type I.

BaMnGaF$_7$ - type II. In this structure [9], half of the Mn$^{2+}$ ions are in distorted square antiprismatic eightfold-coordination. Edge sharing of GaF$_6$ and MnF$_8$ polyhedra occur, forming dinuclear M$_2$F$_{12}$ units (M = Mn, Ga) interconnected by corners to MnF$_6$ octahedra and other M$_2$F$_{12}$ units, building disconnected corrugated layers between which are inserted the Ba atoms. Glass crystallization is the only known way to obtain BaMnFeF$_7$ and BaMnVF$_7$ crystals in that second polymorphic form (type II). So, why a glass with its three-dimensional isotropy, prefers to crystallize in a layered structure rather than into a 3D network of corner-sharing octahedra ? This is unclear.

HT-BaZnFeF$_7$ - type III. This metastable high temperature (prepared at 750°C) structure [10] is characterized by ZnF$_6$ and FeF$_6$ octahedra, and occurence of edge-sharing double groups M$_2$F$_{10}$ of octahedra (M = Zn, Fe), like in type I but differently organized, though both types I and III are forming similar 3D framework structures. Zn$^{2+}$ ions are slightly smaller that Mn$^{2+}$ ones, but this can be adjusted in the starting model. The stable form of BaZnFeF$_7$ is isostructural with the structure type I above.

BaCaGaF$_7$ - type IV. In that structure [11], GaF$_6$ octahedra and CaF$_8$ square antiprisms are linked by corners and edges forming a two dimensional structure. Ga$^{3+}$ and Fe$^{3+}$ ionic radii are similar, the difference between Ca$^{2+}$ and Mn$^{2+}$ (smaller) is not a problem since MnF$_8$ square antiprisms are existing in other structures (type II for instance).

BaCuFeF$_7$ - type V. That structure [12] is characterized by a partial cationic disorder, and is related to HT-BaZnFeF7 (type III), but having a different space group, being built up from edge sharing dioctahedral groups CuFeF$_{10}$ connected by corners in a 3D array.

BaCuInF$_7$ - type VI. In that compound [13], Cu and In are distributed over the same 8c special position in the I4$_1$/amd space group. Using that model for the glass neutron data simulation would not allow to distinguish the (Fe,V)-F and Mn-F distances which are clearly different on the radial distribution functions of the glasses. In order to avoid the M$^{2+}$/M$^{3+}$ disorder, a sub-group (Imma) of the I4$_1$/amd original space group was used. The 3D structure is built up from infinite chains of edge-sharing octahedra. These rutile-like chains are interconnected by octahedra corners.

BaNaZrF$_7$ - type VII. This crystal structure [14] was selected only because of the formula similarity, because Na$^+$ and Zr$^{4+}$ have quite different behaviour from the (Fe/V)$^{3+}$ and Mn$^{2+}$ targets. The 3D structure is build up from infinite zig-zag cis-chains of edge sharing NaF$_8$ cubes linked together by ZrF$_7$ monocapped trigonal prisms. However, it was interesting to see if modelling could adapt that starting model to the neutron data and to different coordinations and interatomic distances constraints. There is systematic microtwinning with this compound. In fact, type VI appears topologically related to types I, III and V.

The crystal data of the seven models are gathered in Table I. Some examples with AMM'F$_7$ compositions supporting the isomorphous Fe/V replacement, are given in Table II. Cell parameters and volume variations are smaller than 5/1000. The crystal structures of BaCd(Fe,V)F$_7$ as well as of BaMn(Fe,V)F$_7$ both belong to the structure-type II. The latter being obtained by crystallizing the title glasses.

Table I - Crystal data for the models with $M^{2+}$ and $M^{3+}$ coordinations.

| Structure-type | *a* | *b* | *c* | *β* | S.G. | V | $\rho_0$ | $M^{2+}$-F | $M^{3+}$-F | Ref. |
|---|---|---|---|---|---|---|---|---|---|---|
| I - BaMnFeF$_7$ | 5.532 | 10.980 | 9.183 | 94.67 | P2$_1$/c | 555.9 | 0.07195 | 6 | 6 | [8] |
| II - BaMnGaF$_7$ | 13.808 | 5.308 | 14.688 | 91.13 | C2/c | 1076.4 | 0.07432 | 6 & 8 | 6 | [9] |
| III - BaZnFeF7 | 5.603 | 9.971 | 9.584 | 92.80 | P2$_1$/c | 560.3 | 0.07139 | 6 | 6 | [10] |
| IV - BaCaGaF7 | 5.390 | 5.410 | 18.978 | 92.3 | P2/n | 552.9 | 0.07235 | 8 | 6 | [11] |
| V - BaCuFeF7 | 10.695 | 9.932 | 5.654 | 118.53 | Cc | 527.7 | 0.07581 | 6 | 6 | [12} |
| VI - BaCuInF7 | 6.843 | 6.843 | 12.001 | 90. | I4$_1$/amd | 522.0 | 0.07663 | 6 | 6 | [13] |
| VII - BaNaZrF$_7$ | 9.118 | 5.5663 | 11.236 | 90. | Pnma | 569.3 | 0.07026 | 8 for Na | 7 for Zr | [14] |

Table II - Some isostructural BaM(Fe/V)F$_7$ crystalline phases.

| Compounds | *a* | *b* | *c* | *β* | S.G. | V | $\rho_0$ | Ref. |
|---|---|---|---|---|---|---|---|---|
| BaCdFeF$_7$ | 13.852 | 5.390 | 15.023 | 91.27 | C2/c | 1121.4 | 0.07134 | [9] |
| BaCdVF$_7$ | 13.853 | 5.400 | 15.043 | 91.50 | C2/c | 1124.9 | 0.07112 | |
| BaMnFeF$_7$ | 13.797 | 5.343 | 14.764 | 90.83 | C2/c | 1088.3 | 0.07351 | this work |
| BaMnVF$_7$ | 13.764 | 5.355 | 14.778 | 90.94 | C2/c | 1089.1 | 0.07345 | |

**RDM modelling**

For the seven structure types, the RDM method was applied by a simultaneous fit of the BaMnFeF$_7$ and BaMnVF$_7$ neutron interference functions. The cell parameters were adjusted to the glass density and then not allowed to vary. The cell parameters and atomic positions used for the structure type II (BaMnGaF$_7$) modelling were those obtained from the refinement of the BaMnFeF$_7$ compound recrystallized from the isoformula glass (Table II). The fit quality by the Rietveld method is characterized by a profile reliability factor, *Rp*, defined as $100*\Sigma|I_{obs}-kI_{calc}|/\Sigma|I_{obs}|$ (%). The reliability factors were calculated for two fit ranges, full range (*Rp$_1$*), and low angle-limited range (*Rp$_2$*), because the full range includes large-angle data which are rather smooth, tending to produce small *Rp* values whatever the fit is good or not. These reliabilities are shown in Table III. According to *Rp$_2$*, the neutron data from the vanadium-based glass (four models with *Rp$_2$* > 10 %) are quite more difficult to reproduce than the data from the iron-based glass (only one model with *Rp$_2$* > 7.5 %). If the model VI (BaCuInF$_7$) can be probably excluded with large *Rp* values, there is no unique choice possible here from the RDM modelling. Models I, II and IV appear equal candidates. The fit for model I is shown in figures 1 (Fe-based glass) and 2 (V-based-glass). It can be seen in fig. 2 that the difficulties in the simulation of the neutron pattern of the V-based glass are coming mainly from the most intense peak at 12°(2θ).

Table III - Rietveld fit conventional reliabilities *Rp$_1$* (%) for the 2.4-124 °2θ full data sets (Fe and V) and a reduced (2.4-44 °2θ) low diffraction angle data set (*Rp$_2$*).

| | I<br>BaMnFeF$_7$ | II<br>BaMnGaF$_7$ | III<br>BaZnFeF$_7$ | IV<br>BaCaGaF$_7$ | V<br>BaCuFeF$_7$ | VI<br>BaCuInF$_7$ | VII<br>BaNaZrF$_7$ |
|---|---|---|---|---|---|---|---|
| *Rp$_1$* (Fe - V) | 2.97 - 3.78 | 3.30 - 3.61 | 3.51 - 4.37 | 3.36 - 4.09 | 3.14 - 4.74 | 6.87 - 6.50 | 3.81 - 5.41 |
| *Rp$_2$* (Fe - V) | 5.90 - 8.62 | 7.15 - 8.86 | 6.99 - 10.36 | 6.89 - 9.24 | 6.55 - 11.30 | 21.2 - 14.7 | 7.41 - 12.14 |

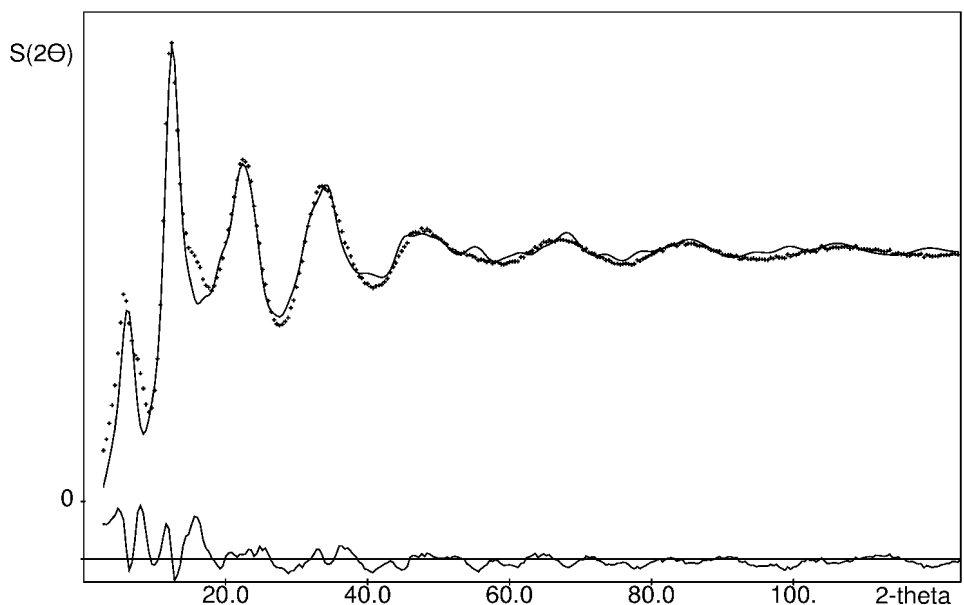

Fig. 1 - Experimental (crosses) and RDM simulated (solid line) neutron data of the BaMnFeF$_7$ glass, using model I. The difference pattern is shown at the bottom.

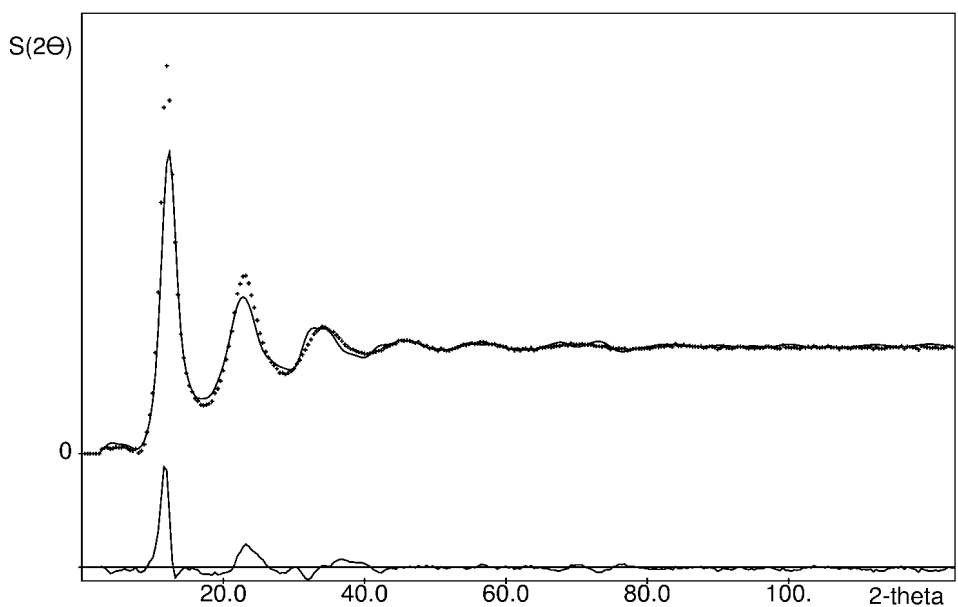

Fig. 2 - Experimental (crosses) and RDM simulated (solid line) neutron data of the BaMnVF$_7$ glass, using model I. The difference pattern is shown at the bottom.

**RMC modelling from the enlarged RDM models**

From the above RDM models, large RMC starting models can be built, and constraints on coordinations and interatomic distance ranges can be applied in order to not destroy the Mn and (Fe,V) polyhedra and their linkage. Table IV shows how were enlarged the starting models and the $Rp$ results. Constraining a part of the same atom-type to have sixfold and the other part eightfold coordination was not that easy. So that, in all cases, only the 6 first F atom neighbours were constrained around Mn and Fe/V atoms. Calculations took several days on fast PCs (processors > 2 GHz) in the range $0.8 < Q < 22.2$ Å$^{-1}$. Considerable improvements in the fit quality is observed when comparing the Tables IV and III. It seems that almost all starting models can lead to small $Rp$ values when applying the RMC method, since $1.91 < Rp_1$

< 2.21 % for the Fe-based glass neutron pattern and 2.50 < $Rp_1$ < 2.67 % for the V-based glass neutron pattern.

Table IV - RMC models built up from RDM ones. Multiplicative factor applied on the a, b, c cell parameters, total number of atoms in the models ($N$), and conventional reliabilities $Rp_1$ (%) for the 0.8-22.2 Q range (Å$^{-1}$) (Fe and V) and a reduced (0.8-9 Å$^{-1}$) low diffraction angle data set ($Rp_2$).

|  | I BaMnFeF$_7$ | II BaMnGaF$_7$ | III BaZnFeF$_7$ | IV BaCaGaF$_7$ | V BaCuFeF$_7$ | VI BaCuInF$_7$ | VII BaNaZrF$_7$ |
|---|---|---|---|---|---|---|---|
| *a* x by | 7 | 3 | 7 | 8 | 4 | 6 | 4 |
| *b* x by | 4 | 6 | 4 | 7 | 4 | 6 | 7 |
| *c* x by | 4 | 3 | 4 | 2 | 7 | 3 | 4 |
| N atoms | 4480 | 4320 | 4480 | 4480 | 4320 | 4320 | 4480 |
| $Rp_1$ (Fe - V) | 2.10 - 2.50 | 1.91 - 2.57 | 1.97 - 2.57 | 2.21 - 2.67 | 2.12 - 2.50 | 2.14 - 2.63 | 2.06 - 2.57 |
| $Rp_2$ (Fe - V) | 4.39 - 5.50 | 4.02 - 5.46 | 4.13 - 5.46 | 4.52 - 5.78 | 4.38 - 5.51 | 4.41 - 5.59 | 4.27 - 5.48 |

If the "best" (defined by the smallest sum of *Rp* values for Fe and V) model appears to be obtained with structure-type II, in which both Fe- and V-based glasses crystallize (BaMnGaF$_7$ structure-type), the difference is quite small. Figures 3 and 4 show the fit obtained with model II. This may mean that the availability of only two interference functions for a four-elements glass (for which ten interference functions would have to be known) gives rise to an undetermined problem, in spite of the coordination constraint and distance restraints. It may also signify that the average orders at short and medium-range distance which characterize all these models are finally very similar, in spite of their obvious differences in the polyhedra modes of connection and in their three- or two-dimensional frameworks. The RDM method operates with a considerably smaller number of degrees of freedom (a few atomic coordinates), and thus produces more clear differences in the fits from various models.

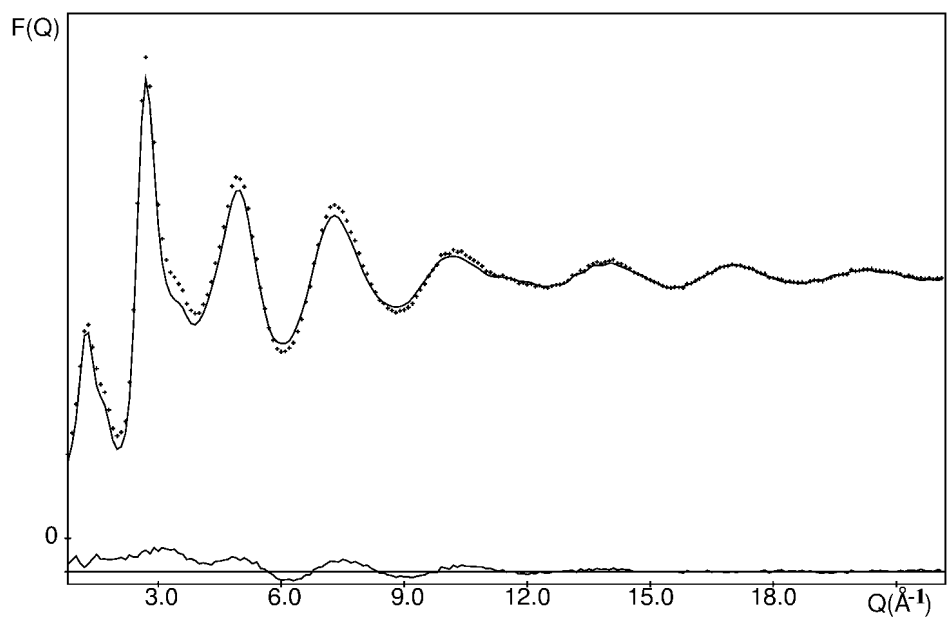

Fig. 3 - Experimental (crosses) and RMC simulated (solid line) neutron data of the BaMnFeF$_7$ glass, using model II. The difference pattern is shown at the bottom.

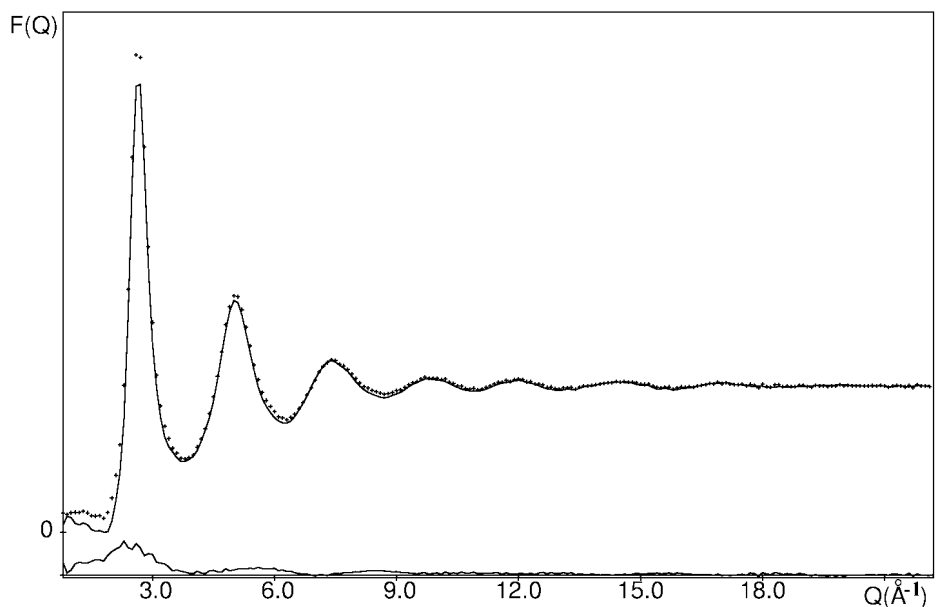

Fig. 4 - Experimental (crosses) and RMC simulated (solid line) neutron data of the BaMnVF$_7$ glass, using model II. The difference pattern is shown at the bottom.

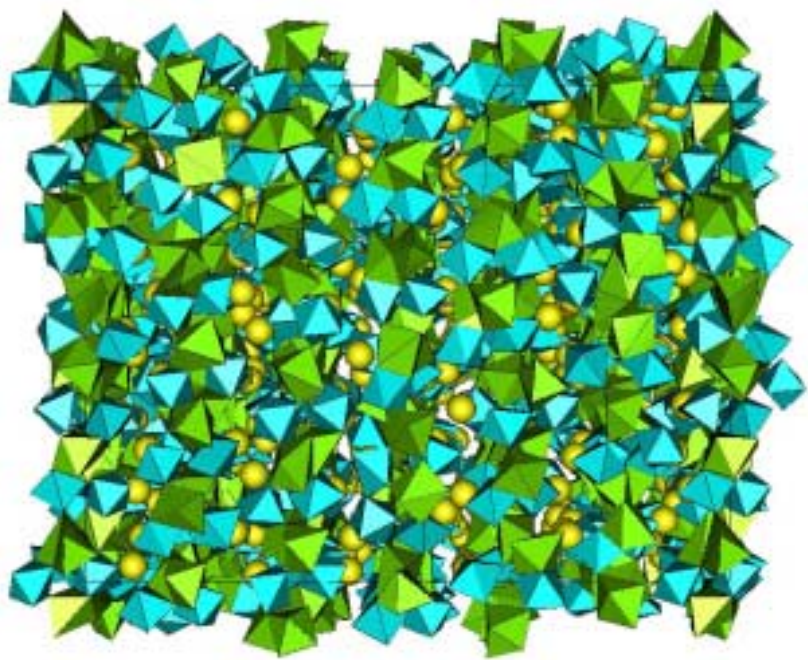

Fig. 5 - The 4320 atoms "best" model II. MnF$_6$ polyhedra in green, (Fe,V)F$_6$ polyhedra in blue, Ba atoms as yellow spheres.

**RMC modelling from random starting models**

Modellings from a fully random starting configurations were processed by using a cubic box of 5000 atoms (cubic edge L = 41.29 Å). Very long runs were needed up to obtain the expected sixfold fluorine coordination around the 3d elements. The "best" of three independent modellings provided $Rp_1$ = 2.11 % (Fe); $Rp_1$ = 2.59 % (V); $Rp_2$ = 4.36 % (Fe); $Rp_2$ = 5.54 % (V) (figures 6 and 7). As previously observed when modelling NaPbFe$_2$F$_9$ glasses [4], the sixfold coordination was fulfilled but the polyhedra show all possibilities between octahedra and trigonal prisms.

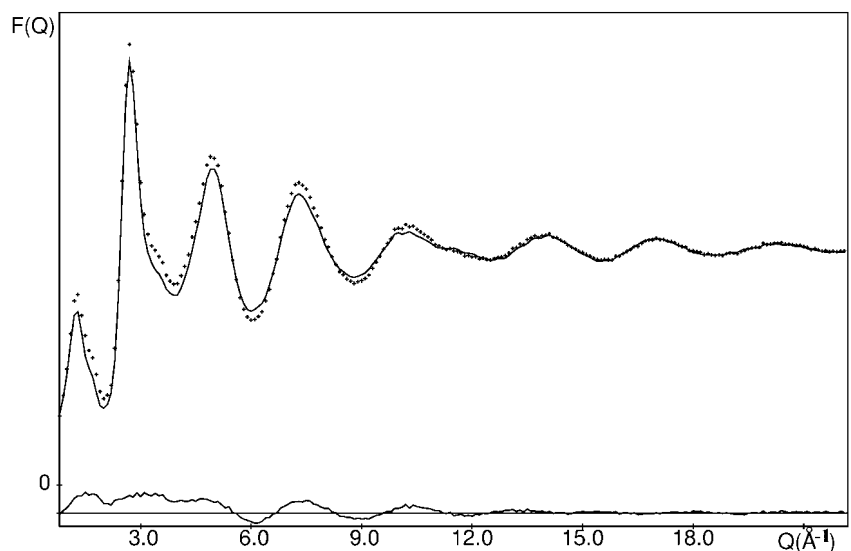

Fig. 6 - Experimental (crosses) and RMC simulated (solid line) neutron data of the $BaMnFeF_7$ glass, starting from a random configuration. The difference pattern is shown at the bottom.

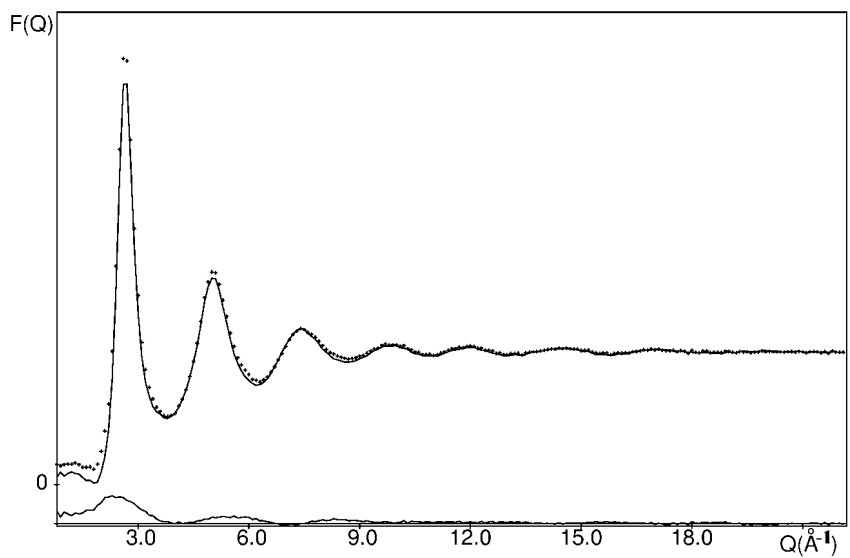

Fig. 7 - Experimental (crosses) and RMC simulated (solid line) neutron data of the $BaMnVF_7$ glass, starting from a random configuration. The difference pattern is shown at the bottom.

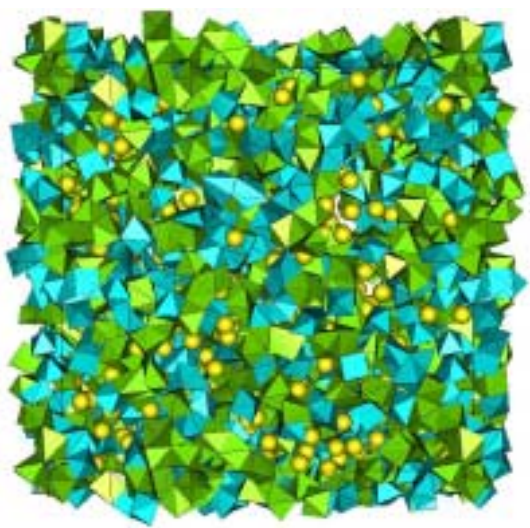

Fig. 8 - The "best" of the 5000 atoms RMC models starting from a random configuration. $MnF_6$ polyhedra in green, $(Fe,V)F_6$ polyhedra in blue, Ba atoms as yellow spheres.

In every cases, the main problems in the agreement between observed and calculated data seems to correspond to a positive difference at $Q < 4$ $\text{Å}^{-1}$. This may be due to some incertitude in the correction from the magnetic form factors contribution of the 3d elements.

**Conclusion**

Several previous consecutive RDM-RMC modelling have shown that the starting crystalline model leading to the most satisfying structure simulation of a glass is generally that of its crystallization product, when it is a unique phase. The conclusion of the present study is again favouring a generalization of this observation since the structure-type II is found to represent the most satisfying model which can be built by RMC among the types I-VII. Even the fully random models built by RMC cannot produce a better agreement between the observed and calculated neutron interference functions. However, there is not a clear gap in fit quality between model II and some others which would allow to claim having elucidated these fluoride glass structures. As usual, the frustrating conclusion is moderated. And the preference of the glass for crystallizing into the structure-type II rather than into the type I is not well understood.

**Acknowledgements**

Many thanks are due to A.-M. Mercier who prepared the glass samples, to the Institut Laue Langevin (Grenoble, France) for providing neutrons (invaluable help in measuring data was ensured by P. Chieux).

# Annex A -

Rietveld crystal structure refinement of BaMnFeF$_7$ in the BaMnGaF$_7$ structure-type, as obtained from the glass recrystallization

```
           ************************************************************
           ** PROGRAM FullProf.2k (Version 2.10 - Mar2002-LLB JRC) **
           ************************************************************
                        M U L T I -- P A T T E R N
                 Rietveld, Profile Matching & Integrated Intensity
                    Refinement of X-ray and/or Neutron Data

     Date: 11/12/2002  Time: 15:59:42.206

 => PCR file code: bmf6
 => DAT file code: bmf6                      -> Relative contribution: 1.0000
 => Title:  BaMnFeF7 glass recrystallizing in the BaMnGaF7 structure type

 ==> CONDITIONS OF THIS RUN FOR PATTERN No.:  1

 => Global Refinement of X-ray powder diffraction data
 => Global Refinement of X-ray powder diffraction data
    Bragg-Brentano or Debye-Scherrer geometry
 => The    5th default profile function was selected

 => Data supplied in free format for pattern:  1
 => Wavelengths:  1.54056 1.54439
 => Cos(Monochromator angle)=   0.7998
 => Absorption correction (AC), muR-eff =    0.0000
 => Base of peaks: 2.0*HW*   12.00
 ==> Angular range, step and number of points:
     2Thmin:  10.0000 2Thmax:  79.9800 Step:   0.0200 No. of points:    3500
 => STRUPLO File (*.sch) is output for phase: 1
 =>-------> Pattern#  1
 => Crystal Structure Refinement for phase: 1
 => Scor: 3.7826

 ==> RESULTS OF REFINEMENT:

 => No. of fitted parameters:   42

--------------------------------------------------------------------------------
 => Phase No.  1   "BaMnFeF7"                                 C 2/C
--------------------------------------------------------------------------------

 =>  No. of reflections for pattern#:   1:   774/2

 ==> ATOM PARAMETERS:

  Name      x       sx       y        sy       z       sz       B       sB   occ.   socc.  Mult

 BA    0.19038(  11) -0.04375(  30)  0.12082(  12)  2.026( 44)  0.040(  0)    8
 MN1   0.00000(   0)  0.46147( 116)  0.25000(   0)  1.006( 66)  0.020(  0)    4
 MN2   0.00000(   0)  0.50000(   0)  0.00000(   0)  1.006( 66)  0.020(  0)    4
 FE    0.37942(  24)  0.49199(  89)  0.12271(  28)  1.006( 66)  0.040(  0)    8
 F1    0.16659(  92)  0.19586( 189)  0.27574(  86)  1.098(107)  0.040(  0)    8
 F2    0.41818(  95)  0.33326( 185)  0.01457(  80)  1.098(107)  0.040(  0)    8
 F3    0.37781(  94)  0.23027( 211)  0.21386(  93)  1.098(107)  0.040(  0)    8
 F4    0.37373(  84)  0.21719( 213)  0.54999(  93)  1.098(107)  0.040(  0)    8
 F5    0.51166(  75)  0.36373( 170)  0.65946(  68)  1.098(107)  0.040(  0)    8
 F6    0.24529(  74)  0.44748( 193)  0.09144(  66)  1.098(107)  0.040(  0)    8
```

```
 F7    0.44665(  79)  0.07233( 179)  0.36929(  85)   1.098(107)   0.040(  0)       8

 ==> PROFILE PARAMETERS FOR PATTERN#   1

 => Cell parameters        :
                                13.79730    0.00083
                                 5.34319    0.00031
                                14.76393    0.00083
                                90.00000    0.00000
                                90.83080    0.00232
                                90.00000    0.00000

 => overall scale factor :    0.441444546    0.001607108
 => Eta(p-v) or m(p-vii) :    0.64707    0.00986
 => Overall tem. factor  :    0.00000    0.00000
 => Halfwidth parameters :    0.13792    0.00533
                             -0.00805    0.00000
                              0.02685    0.00056
 => Preferred orientation:    1.00000    0.00000
                              0.00000    0.00000
 => Asymmetry parameters :    0.02800    0.01359
                              0.02498    0.00123
                              0.00000    0.00000
                              0.00000    0.00000
 => X and y parameters   :    0.00000    0.00000
                              0.00000    0.00000
 => Strain parameters    :    0.00000    0.00000
                              0.00000    0.00000
                              0.00000    0.00000
 => Size parameters (G,L):    0.00000    0.00000
                              0.00000    0.00000

 ==> GLOBAL PARAMETERS FOR PATTERN#   1

 => Zero-point:     0.0456    0.0046
 => Cos( theta)-shift parameter :   0.0000  0.0000
 => Sin(2theta)-shift parameter :   0.0000  0.0000

 ==> RELIABILITY FACTORS WITH ALL NON-EXCLUDED POINTS FOR PATTERN:   1

 => Cycle: 10 => MaxCycle: 10
 => N-P+C:  3309
 =>  R-factors (not corrected for background) for Pattern:  1
 => Rp: 5.22     Rwp: 6.95     Rexp:    1.44 Chi2:  23.4       L.S. refinement
 => Conventional Rietveld R-factors for Pattern:  1
 => Rp: 9.27     Rwp: 10.8     Rexp:    2.23 Chi2:  23.4
 => Deviance: 0.767E+05     Dev*  :   23.18
 => DW-Stat.:    0.5354     DW-exp:     1.9180
 => N-sigma of the GoF:  911.995

 ==> RELIABILITY FACTORS FOR POINTS WITH BRAGG CONTRIBUTIONS FOR PATTERN:   1

 => N-P+C:  3280
 =>  R-factors (not corrected for background) for Pattern:  1
 => Rp: 5.23     Rwp: 6.96     Rexp:    1.43 Chi2:  23.6       L.S. refinement
 => Conventional Rietveld R-factors for Pattern:  1
 => Rp: 9.23     Rwp: 10.8     Rexp:    2.22 Chi2:  23.6
 => Deviance: 0.766E+05     Dev*  :   23.34
 => DW-Stat.:    0.5365     DW-exp:     1.9178
 => N-sigma of the GoF:      914.403

 => Global user-weigthed Chi2 (Bragg contrib.):   23.6
 => Phase:  1
 => Bragg R-factor:   5.51         Vol: 1088.305( 0.110)  Fract(%):  100.00( 0.52)
 => Rf-factor= 3.88             ATZ:          1180.170   Brindley:  1.0000
```

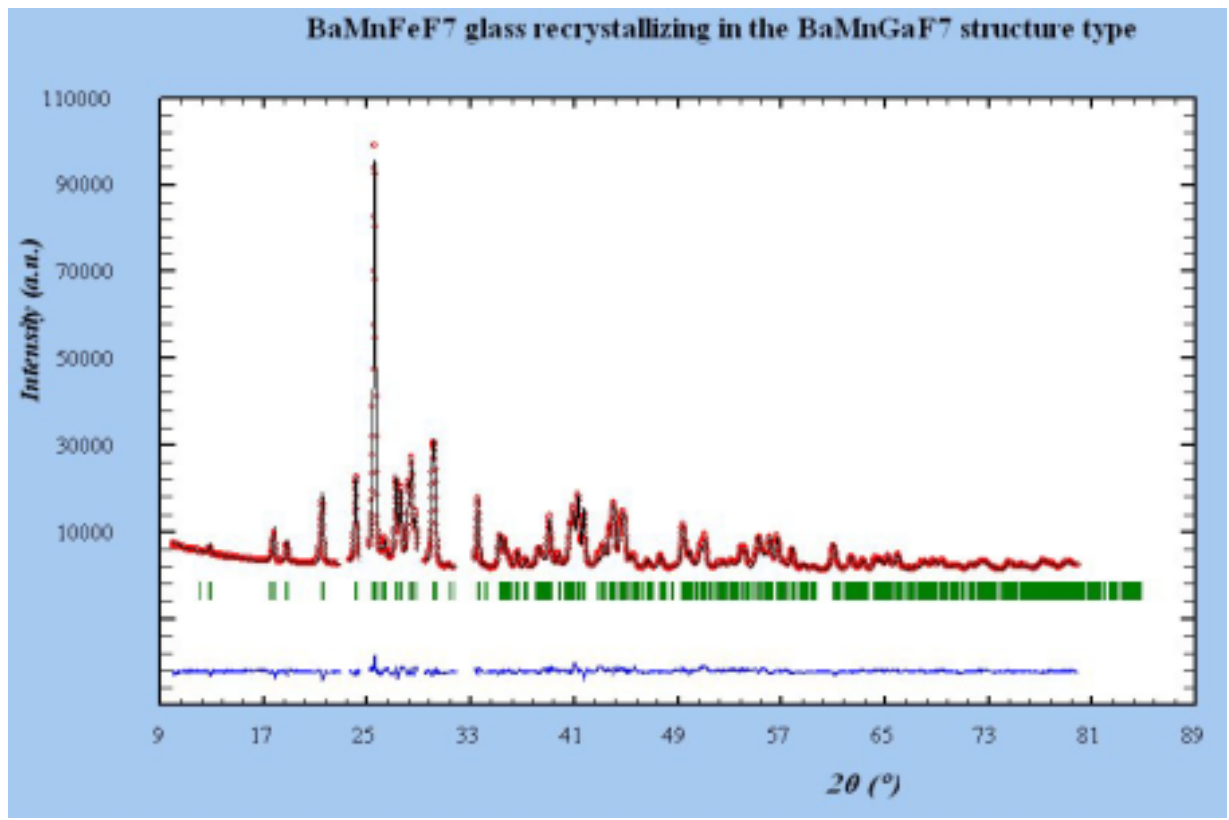

**Plot of the Rietveld fit**